\documentclass[twocolumn,prl,showpacs]{revtex4}
\usepackage{amssymb}
\usepackage{amsmath}
\usepackage{graphicx}

\begin{document}

\title{ Decay of Loschmidt Echo Enhanced by Quantum Criticality}
\author{H.T. Quan$^{1}$, Z. Song$^{2}$, X.F. Liu $^{3}$, P. Zanardi $^{4}$
and C.P. Sun$^{1,2}$ }
\email{suncp@itp.ac.cn}
\homepage{http://www.itp.ac.cn/~suncp}
\affiliation{$^{1}$ Institute of Theoretical Physics, Chinese Academy of Sciences,
Beijing, 100080, China}
\affiliation{$^{2}$Department of Physics, Nankai University, Tianjin 300071, China}
\affiliation{$^{3}$ Department of Mathematics, Peking University, Beijing, 100871, China}
\affiliation{$^{4}$ Institute for Scientific Interchange Foundation Villa Gualino Viale
Settimio Severo 65 I-10133 Torino, Italy}

\begin{abstract}
We study the transition of a quantum system $S $ from a pure state to a
mixed one, which is induced by the quantum criticality of the surrounding
system $E$ coupled to it. To characterize this transition quantitatively, we
carefully examine the behavior of the Loschmidt echo (LE) of $E$ modelled as
an Ising model in a transverse field, which behaves as a measuring apparatus
in quantum measurement. It is found that the quantum critical behavior of $E$
strongly affects its capability of enhancing the decay of LE: near the
critical value of the transverse field entailing the happening of quantum
phase transition, the off-diagonal elements of the reduced density matrix
describing $S$ vanish sharply.
\end{abstract}

\pacs{03.65.Yz, 05.70.Jk, 03.65.Ta, 05.50.+q, }
\maketitle

\emph{Introduction:} Nowadays quantum -classical transitions described by a
reduction from a pure state to a mixture \cite{Deco,Zurek} renew interests
in many areas of physics, mainly due to the importance of quantum
measurement and decoherence problem in quantum computing. To study this
transition, some exactly-solvable models were proposed for a system coupled
to the macroscopic \cite{Hepp,Bell, namik} or classical \cite{Cini,Sun1}
surrounding systems. Relevantly, in association with the quantum -classical
transition in quantum chaos, the concept of Loschmidt echo (LE) from NMR
experiments was introduced to describe the hypersensitivity of the time
evolution to the perturbations experienced by the surrounding system\cite%
{peres,q-chaos}. In this letter, by a concrete example, we will show how
quantum phase transition (QPT) \cite{Sachdev} of the surrounding system can
also sensitively affect the decay of its own LE, which means a dynamic
reduction of its coupled system from pure state to a mixed one. Here, we
note that a QPT effect has been explored for the Dicke model at the
transition from quasi-integrable to quantum chaotic phases \cite{Dicke}.

As a quantum critical phenomenon, QPT happens at zero temperature, at which
the thermal fluctuations vanish. Thus QPT is driven only by quantum
fluctuation, and the uncertainty relation lie at the heart of various QPT
phenomena. On the other hand, the randomness of the relative phase, which
causes pure-mixed state transition, also has its source in the uncertainty
principle \cite{ZLS}. It is this observation that enlightens us to explore
the relationship between QPT and the pure-mixed state transition described
dynamically by the time evolution of the LE. It is common that the ground
state of the critical system is very sensitive to the varying magnitude of
the coupling constant \cite{Zurek1}, or the system experiences a spontaneous
symmetry breaking at the critical point. Up to now, all of the known models
of QPT possess this property. Actually, this kind of critical sensitiveness
can be well understood by resorting to the concepts of quantum chaos through
the LE \cite{q-chaos} or macroscopic enhancement of phase randomness \cite%
{ZLS}.

Our approach is based on the Hepp-Coleman (HC) model \cite{Hepp,Bell}, which
was initially proposed as a model for quantum measurement. In our
generalization, the free spin 1/2 ensemble, as a model of measuring
apparatus, is replaced by the Ising spin chain $E$ in a transverse field,
and the two-level central system $S$ interacts with this spin chain
transversely \cite{Sachdev,Pfeuty}. Corresponding to the two basis vectors
of $S$, the interaction between $E$ and $S$ then leads to two slightly
different effective Hamiltonians acting on $E$. The crucial point is that
these two effective Hamiltonians have distinguished ground state symmetries
near the critical point. This is just what underlies the decay of LE induced
by the criticality of the surrounding system $E$.

\begin{figure}[h]
\begin{center}
\includegraphics[bb=107 275 495 545, width=7cm, clip]{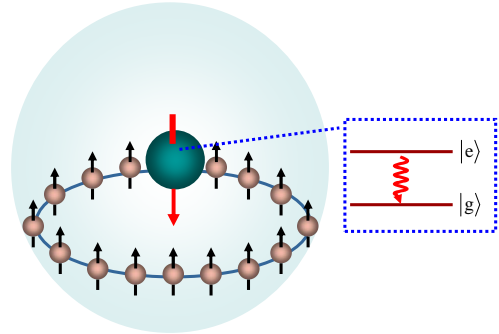}
\end{center}
\caption{(Color online) A schematic diagram of the physical
implementation of the generalized Hepp-Coleman Model. The spins
are arranged in a circle to form a ring array $E$. The central
two-level system $S$ possesses homogeneous couplings due to the
overlaps of symmetric spacial wave function of $S$ with those of
spins.}
\end{figure}

\emph{Model setup based on quantum phase transition: } Our model concerning
the decay of LE, illustrated in Fig. 1, is very similar to the Hepp-Coleman
model \cite{Hepp, Bell} or its generalizations \cite{Cini, namik,Sun1}. We
take the surrounding system $E$ to be an Ising spin chain in a transverse
field, which satisfies the Born-Von Karman condition automatically. Consider
a two level system $S$ with the excited state $\left\vert e\right\rangle $
and the ground state $\left\vert g\right\rangle $, which is transversely
coupled to $E$. The corresponding Hamiltonian reads as follows:%
\begin{equation}
H(\lambda ,\delta )=-J\sum\limits_{j}\left( \sigma _{j}^{z}\sigma
_{j+1}^{z}+\lambda \sigma _{j}^{x}+\delta \left\vert e\right\rangle
\left\langle e\right\vert \sigma _{j}^{x}\right) ,  \label{1}
\end{equation}%
where $J$ and $\lambda $\ characterize the strengths of the Ising
interaction and the coupling to the transverse field respectively; $\delta $
indicates the small perturbation coupling of $S$ to $E$; $\sigma
_{i}^{\alpha }$ ($\alpha =x,y,z$) are the Pauli operators defined on the $i$%
th site in the lattice with lattice spacing $a$.

We assume the two-level system is initially in a superposition state $%
\left\vert \phi _{s}\left( 0\right) \right\rangle =c_{g}\left\vert
g\right\rangle +c_{e}\left\vert e\right\rangle $, where the coefficients $%
c_{g}$ and $c_{e}$ satisfy $\left\vert c_{g}\right\vert ^{2}+\left\vert
c_{e}\right\vert ^{2}=1$. Then the evolution of the Ising spin chain
initially prepared in $\left\vert \varphi \left( 0\right) \right\rangle $,
will split into two branches $\left\vert \varphi _{\alpha }\left( t\right)
\right\rangle =\exp (-iH_{\alpha }t)\left\vert \varphi \left( 0\right)
\right\rangle $ $\left( \alpha =e,g\right) $, and the total wave function is
obtained as $\left\vert \psi \left( t\right) \right\rangle =c_{g}\left\vert
g\right\rangle \otimes \left\vert \varphi _{g}\left( t\right) \right\rangle
+c_{e}\left\vert e\right\rangle \otimes \left\vert \varphi _{e}\left(
t\right) \right\rangle $. Here, the evolutions of the two branch wave
functions $\left\vert \varphi _{\alpha }\left( t\right) \right\rangle $ are
driven respectively by the two effective Hamiltonians $H_{g}=H(\lambda ,0)$
and $H_{e}=H(\lambda ,\delta )\equiv H_{g}+V_{e}$. Obviously, both $H_{g}$\
and $H_{e}$\ describe the Ising model in a transverse field, but with a tiny
difference in the field strength. The central quantum system in two
different states $\left\vert e\right\rangle $ and $\left\vert g\right\rangle
$ will exert slightly different back actions on the surrounding system,
which manifest as two effective potentials $V_{e}=-J\delta \sum_{j}\sigma
_{j}^{x}$ and $V_{g}=0$. This difference just results in the decay of the LE
\cite{q-chaos} defined as
\begin{equation}
L(\lambda ,t)=|\langle \varphi _{g}\left( t\right) \left\vert \varphi
_{e}\left( t\right) \right\rangle |^{2}.  \label{3}
\end{equation}

To show the key role of LE in depicting quantum decoherence of the central
system, and manifest the difference between LE and decoherence, we define
the purity \cite{purity} $P=Tr_{S}\left( \rho _{S}^{2}\right)
=Tr_{S}\{[Tr_{E}\rho \left( t\right) ]^{2}\}$\ to describe the decoherence.
Here, $\rho \left( t\right) =\left\vert \psi \left( t\right) \right\rangle
\left\langle \psi \left( t\right) \right\vert $, and $Tr_{\alpha }$\ means
tracing over the variable of $\alpha $, $\left( \alpha =E,S\right) $. A
straightforward calculation reveals the relationship between LE and the
purity as $P=1-2\left\vert c_{e}c_{g}\right\vert ^{2}\left[ 1-L(\lambda ,t)%
\right] $. This equation indicates that the purity depends on the initial
states of the central system and the surrounding system $E$, but the LE only
depends on the initial state of $E$. For simplicity, we assume the
surrounding system is initially prepared in the ground state. In the
following discussion, we will focus on the decay problem of LE induced by
the coupling with the central system.

\emph{Exact solution for the Loschmidt echo: } We now prepare to prove that,
just at the critical point $\lambda =\lambda _{c}=1$, the decay of LE is
enhanced, accompanied by the QPT in one of the two evolution branches.

We first diagonalize the effective Hamiltonian as $H_{e}=\sum_{k}\varepsilon
_{e}^{k}\left( A_{k}^{\dagger }A_{k}-1/2\right) $ in terms of the normal
mode operators \cite{Sachdev, Pfeuty}
\begin{equation}
A_{k}=\sum_{l}\frac{e^{-ikal}}{\sqrt{N}}\prod\limits_{s<l}\sigma _{s}^{\left[
x\right] }\left( u_{e}^{k}\sigma _{l}^{\left[ +\right] }-iv_{e}^{k}\sigma
_{l}^{\left[ -\right] }\right) ,  \label{4}
\end{equation}%
which satisfy the canonical fermion anti-commutation relations. Here, $N$ is
the number of sites of the spin chain, and $\sigma _{l}^{\left[ \pm \right]
}=\left( -\sigma _{l}^{z}\pm i\sigma _{l}^{y}\right) /2$ is defined by the
Pauli matrices $\sigma _{l}^{\alpha },\alpha =x,y,z$. The\textbf{\ }%
coefficients $u_{e}^{k}=\cos \left( \theta _{e}^{k}/2\right) ,v_{e}^{k}=\sin
\left( \theta _{e}^{k}/2\right) $ depend on the the angle
\begin{equation}
\theta _{e}^{k}=\theta _{e}^{k}(\delta )=\arctan [\frac{-\sin \left(
ka\right) }{\cos \left( ka\right) -\left( \lambda +\delta \right) }]
\end{equation}%
The corresponding single quasi-excitation energy $\varepsilon _{e}^{k}$ is%
\begin{equation}
\varepsilon _{e}^{k}(\delta )=2J\sqrt{1+\left( \lambda +\delta \right)
^{2}-2\left( \lambda +\delta \right) \cos \left( ka\right) }.
\end{equation}
Note that, in writing down the known result (\ref{4}) in a compact form, we
have combined the Jordan-Wigner map and the Fourier transformation to the
momentum space \cite{Sachdev, Pfeuty}.

The effective Hamiltonian $H_{g}$ can be diagonalized in a similar way: $%
H_{g}=\sum_{k}\varepsilon _{g}^{k}\left( B_{k}^{\dagger }B_{k}-1/2\right) $.
In this case the single quasi-excitation energy is $\varepsilon
_{g}^{k}=\varepsilon _{e}^{k}(0)$ and the corresponding fermionic
quasi-excitation operators $B_{k}$ can be obtained by the following
Bogliubov transformation%
\begin{equation}
B_{\pm k}=\cos \left( \alpha _{k}\right) A_{\pm k}-i\sin \left( \alpha
_{k}\right) \left( A_{\mp k}\right) ^{\dagger }.  \label{6}
\end{equation}%
Here, $\alpha _{k}=[\theta _{g}^{k}-\theta _{e}^{k}]/2$, and $\theta
_{g}^{k} $ are defined by $\theta _{g}^{k}=\theta _{e}^{k}(0).$

We suppose that the spin chain is initially in the ground state $\left\vert
\varphi \left( 0\right) \right\rangle =\left\vert G\right\rangle _{g}$ of
the Ising spin chain in a transverse field depicted by $H_{g}$, i.e., $%
B_{k}\left\vert G\right\rangle _{g}=0$ for any operator $B_{k}$. Then from
Eq. (\ref{6}) the state $\left\vert G\right\rangle _{g}$ can be rewritten as
a BCS-like state:
\begin{equation}
\left\vert G\right\rangle _{g}=\prod\limits_{k>0}\left[ \cos \left( \alpha
_{k}\right) -i\sin \left( \alpha _{k}\right) A_{k}^{\dagger }A_{-k}^{\dagger
}\right] \left\vert G\right\rangle _{e},  \label{7}
\end{equation}%
where $\left\vert G\right\rangle _{e}$ is the ground state of $H_{e}$. This
explicit expression of $\left\vert G\right\rangle _{g}$ enables us to
calculate straightforwardly the LE (\ref{3}), which assumes the following
factorized form:%
\begin{equation}
L(\lambda ,t)=\prod\limits_{k>0}F_{k}=\prod\limits_{k>0}[1-\sin ^{2}\left(
2\alpha _{k}\right) \sin ^{2}\left( \varepsilon _{e}^{k}t\right) ].
\label{8}
\end{equation}

\emph{Quantum-classical transition at critical point of QPT: }Since each
factor $F_{k}$ in Eq (\ref{8}) has a norm less than unity, we may well
expect $L(\lambda ,t)$ to decrease to zero in the large $N$ limit under some
reasonable conditions. This kind of factorized structure was first
discovered and systematically studied by one of the authors in developing
the quantum measurement theory in classical or macroscopic limit \cite{Sun1}
and it has been applied to analyze the universality of decoherence influence
from environment on quantum computing \cite{SZL}. Now we study in detail the
critical behavior of the surrounding system near the critical point $\lambda
_{c}=1$ and its relation to the sensitive evolution of the LE perturbed by
the central system even for a finite $N$. This turns out to reveal a novel
mechanism responsible for the enhanced decay of LE.

Let us first make a heuristic analysis of the features of the LE. For a
cut-off frequency $K_{c}$ we define the partial product for the LE
\begin{equation}
L_{c}(\lambda ,t)\equiv \prod\limits_{k>0}^{K_{c}}F_{k}\emph{\ }\geq
L(\lambda ,t),  \label{9}
\end{equation}%
and the corresponding partial sum $S(\lambda ,t)=\ln L_{c}\equiv
-\sum_{k>0}^{K_{c}}|\ln F_{k}|$. For small $k$\ we have $\varepsilon
_{e}^{k}\approx 2J|1-\lambda -\delta |$, $\sin ^{2}\left[ 2\alpha _{k}\right]
\approx \left( \delta ka\right) ^{2}/(1-\lambda )^{2}\left( 1-\lambda
-\delta \right) ^{2}$. As a result, if $K_{c}$\ is small enough we have%
\begin{equation}
S(\lambda ,t)\approx -\frac{\delta ^{2}E(K_{c})\sin ^{2}\left( 2Jt\left\vert
1-\lambda -\delta \right\vert \right) }{(1-\lambda )^{2}\left( 1-\lambda
-\delta \right) ^{2}},
\end{equation}%
where $E(K_{c})=4\pi ^{2}N_{c}(N_{c}+1)(2N_{c}+1)/(6N^{2})$\ and $N_{c}$\ is
the integer nearest to $NK_{c}a/2\pi $.\textbf{\ }Here we have used the fact
that the Bloch wave vector $k$ takes the discrete values $2n\pi /Na$ $\left(
n=1,2,\cdots N/2\right) $. In this case, it then follows that for a fixed $%
t, $
\begin{equation}
L_{c}(\lambda ,t)\approx \exp \left( -\gamma t^{2}\right)  \label{11}
\end{equation}%
when $\lambda \rightarrow \lambda _{c}=1$,\ where $\gamma =4J^{2}\delta
^{2}E(K_{c})/(1-\lambda )^{2}.$

Notice that $L(\lambda ,t)$ is less than $L_{c}(\lambda ,t)$. So from the
above heuristic analysis we may expect that, when $N$ is large enough and $%
\lambda $ is adjusted to the vicinity of the critical point $\lambda _{c}=1$%
, the LE will exceptionally vanish with time. On the other hand, we observe
that $\gamma $ seems to approach zero in the thermodynamic limit $%
N\rightarrow \infty $ for $Na$ keeps as a constant in the process of taking
this limit and $E(K_{c})\propto 1/N^{2}$. Since a true QPT can occur just in
the thermodynamic limit, it is natural to doubt whether the QPT, and thus
the induced decay of the LE, can happen at the critical point. In fact, due
to the vanishing denominator $(1-\lambda )^{2}$ of $\gamma $ at the critical
point of the QPT, the decay of the LE is still possible even for $\gamma $
having a vanishing numerator. For a practical system used to demonstrate the
QPT induced decay of the LE, the particle number $N$ is large, but finite,
and then the practical $\gamma $ does not vanish.

Now we resort to numerical calculation to test the heuristic analysis. For $%
N=50\sim 250$\textbf{, }$\delta =0.1$\textbf{, }the LE are calculated
numerically from the exact expression (\ref{8}) with the parameters within
the ranges $\lambda \in \lbrack 0,2]$, $t\in \lbrack 0,27/J]$. The results
are demonstrated in Figs. 2a and 2b.
\begin{figure}[tbp]
\includegraphics[bb=119 274 488 634, width=4 cm, clip]{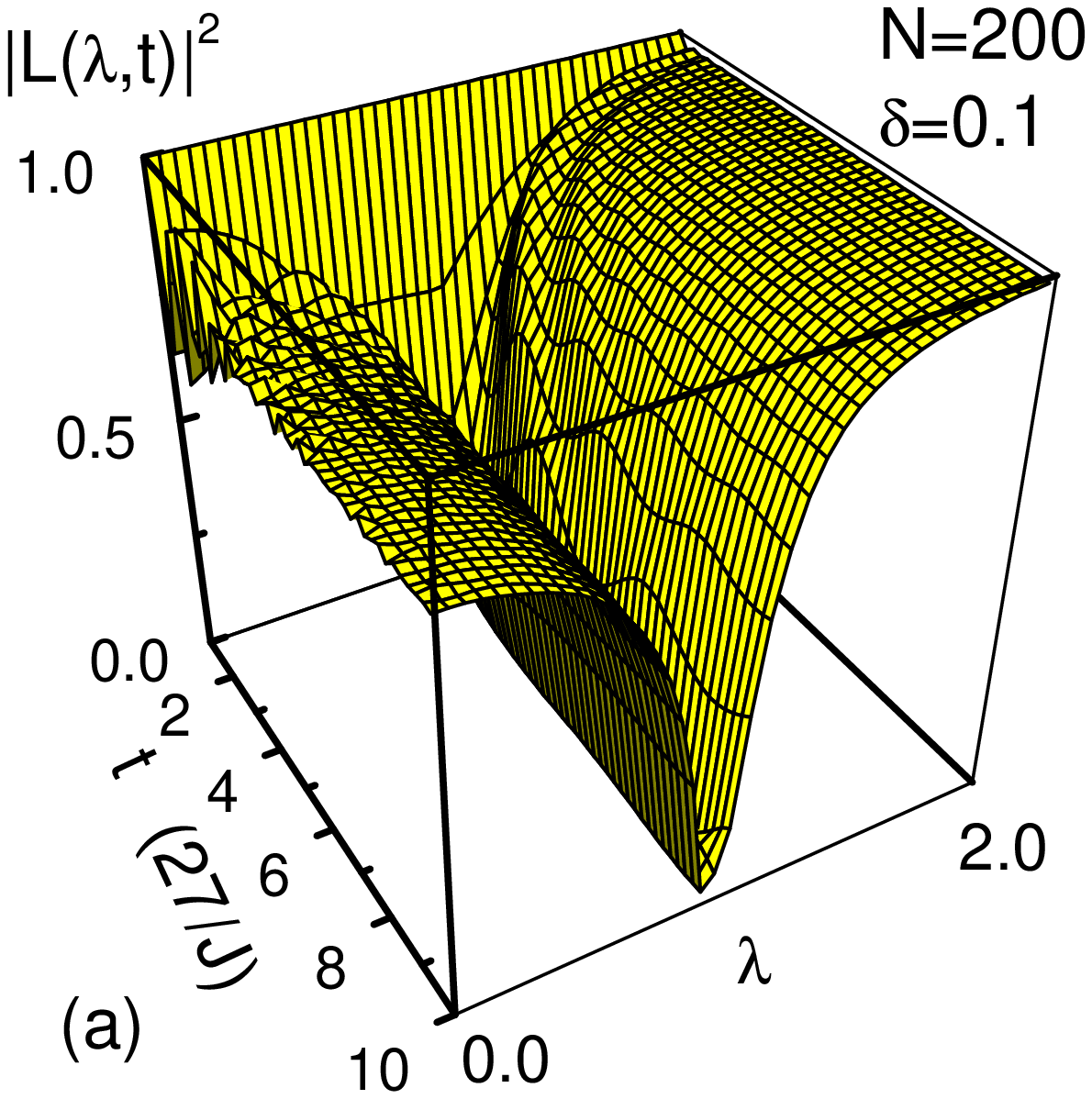} %
\includegraphics[bb=131 156 447 473, width=4 cm, clip]{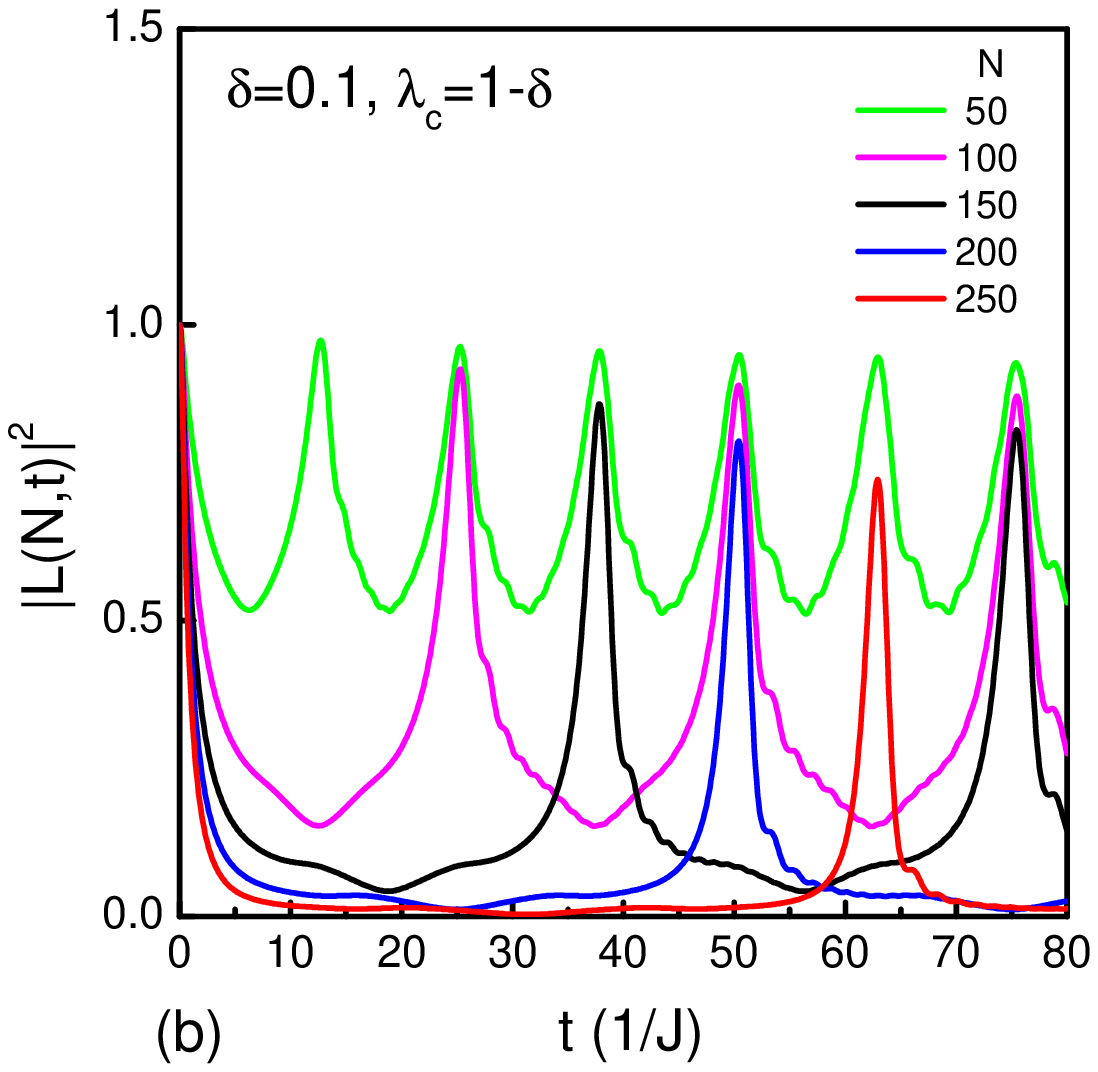}
\caption{(Color online) (a) Three dimensional (3-D) diagram of the LE $%
|L\left( \protect\lambda ,t\right) |^{2}$ as the function of $\protect%
\lambda $ and $t$ for the system with $N=200$. The valley around the
critical point $\protect\lambda _{c}$=$1$ indicates that the decay of LE is
enhanced by the QPT . The profile at \textbf{$\protect\lambda =1$} is in
agreement with the analytical analysis. (b) The cross sections of the 3-D
surface for the systems of $N=50,100,150,200,$ and $250$ at $\protect\lambda
$= $\protect\lambda _{c}$-$\protect\delta $=$0.9$. It shows that the
quasi-period of the LE is proportional to the size of the surrounding system
.}
\end{figure}

In Fig. 2a there exists a deep valley in the domain around the line $\lambda
=\lambda _{c}-\delta =0.9$. This reflects the fact that near the critical
point of the surrounding system the LE is very sensitive to the perturbation
experienced by the surrounding system. At the critical point, with a highly
enhanced decay of LE, the central system transits from a pure state to a
mixed state due to its entanglement with the surrounding system. The five
curves in Figure 2b clearly demonstrate the influence of $N$ on the decay
behavior of the LE. At $\lambda =\lambda _{c}-\delta =0.9$, the LE decays
and revives as time increases. The period of the revival of the LE is
proportional to the size of the surrounding system.

\emph{Decays and revivals of Loschmidt echo as a witness of QPT: } The novel
phenomenon of the synchronization between the QPT and the enhanced decay of
the LE mentioned above and its physical implication deserves further
exploring. Generally, the two terms in $H_{e}$ represent two competitive
physical effects with different order tendencies: in the weak coupling case $%
\lambda \ll 1$ the ground state is either all spins up or all spins down,
while in the strong coupling case $\lambda \gg 1$ the ground state tends to
the saturated ferromagnetic state with all the spins pointing right. When $%
\lambda $\ takes the value of the order unity, the qualitative properties of
the ground states for $\lambda >1$ and $\lambda <1$ are similar to those for
$\lambda \gg 1$ and $\lambda \ll 1$ respectively. Only the critical point $%
\lambda =1$ has genuinely different properties.

The singular behavior of QPT at $\lambda =\lambda _{c}$\ reflects the
hypersensitivity of the ground states of the surrounding system with respect
to the perturbation coupling imposed by the central system, which is
reflected by the evolution of LE. We can thus expect quantum evolution of
the surrounding system to inherit this sensitivity, which can also be
understood as a signature of quantum chaos: For a quantum system prepared in
the identical initial state, two slightly different interactions can lead to
two quite different quantum evolutions. Mathematically speaking, this means
the LE, initially equal to $1$, will decay with time and finally vanish. In
this sense the sensitivity of quantum evolution to perturbation plays a
crucial role in inducing the decay of LE. Due to the perturbations of two
effective potentials caused by $\left\vert e\right\rangle $ and $\left\vert
g\right\rangle $ respectively, the LE can decrease to zero due to the
singularity at the critical point and the macroscopic enhancement of phase
randomness for large $N$, only at which QPT occurs \cite{ZLS}.

\begin{figure}[tbp]
\includegraphics[bb=119 274 488 634, width=4 cm, clip]{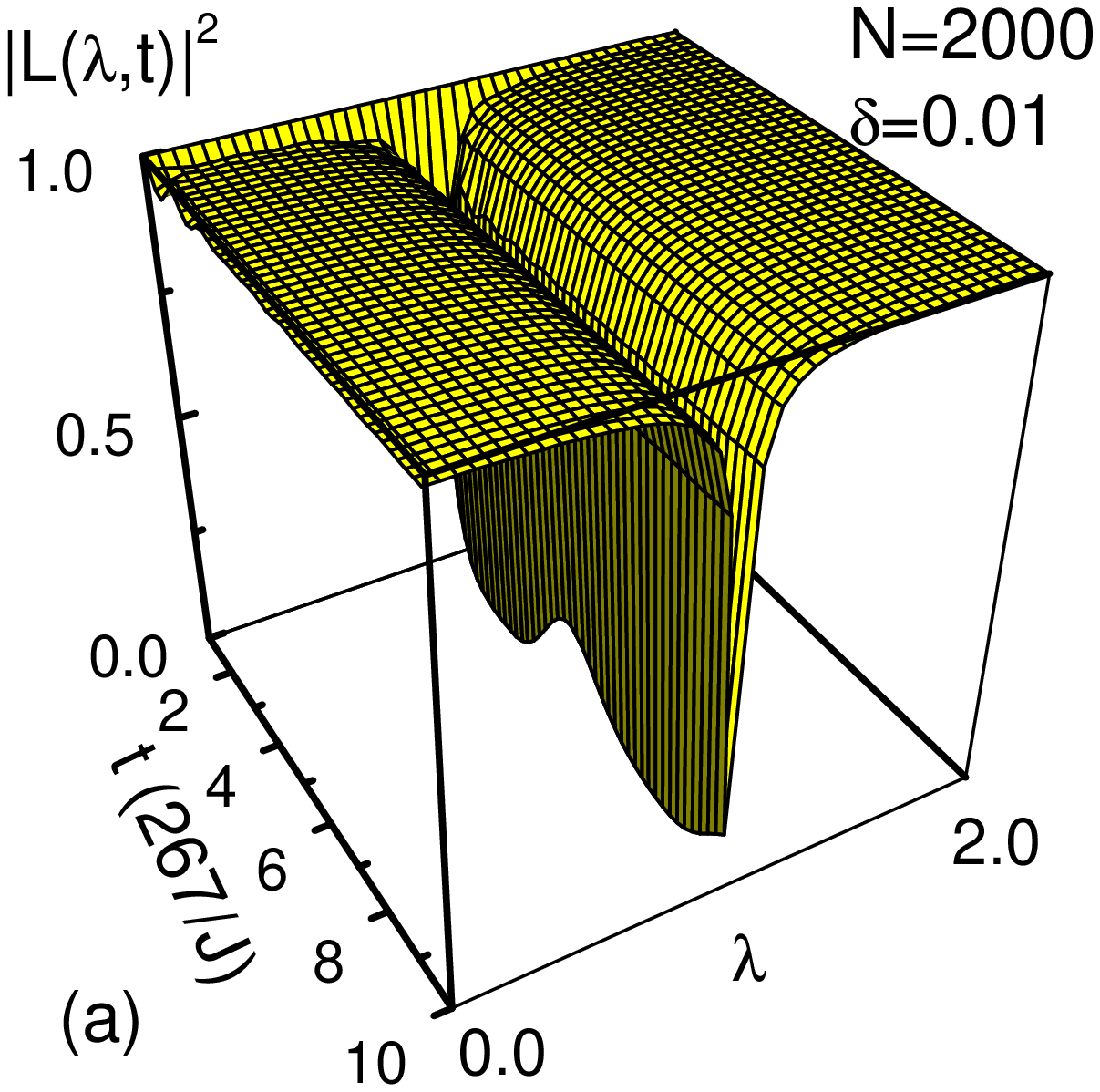} %
\includegraphics[bb=131 156 447 473, width=4 cm, clip]{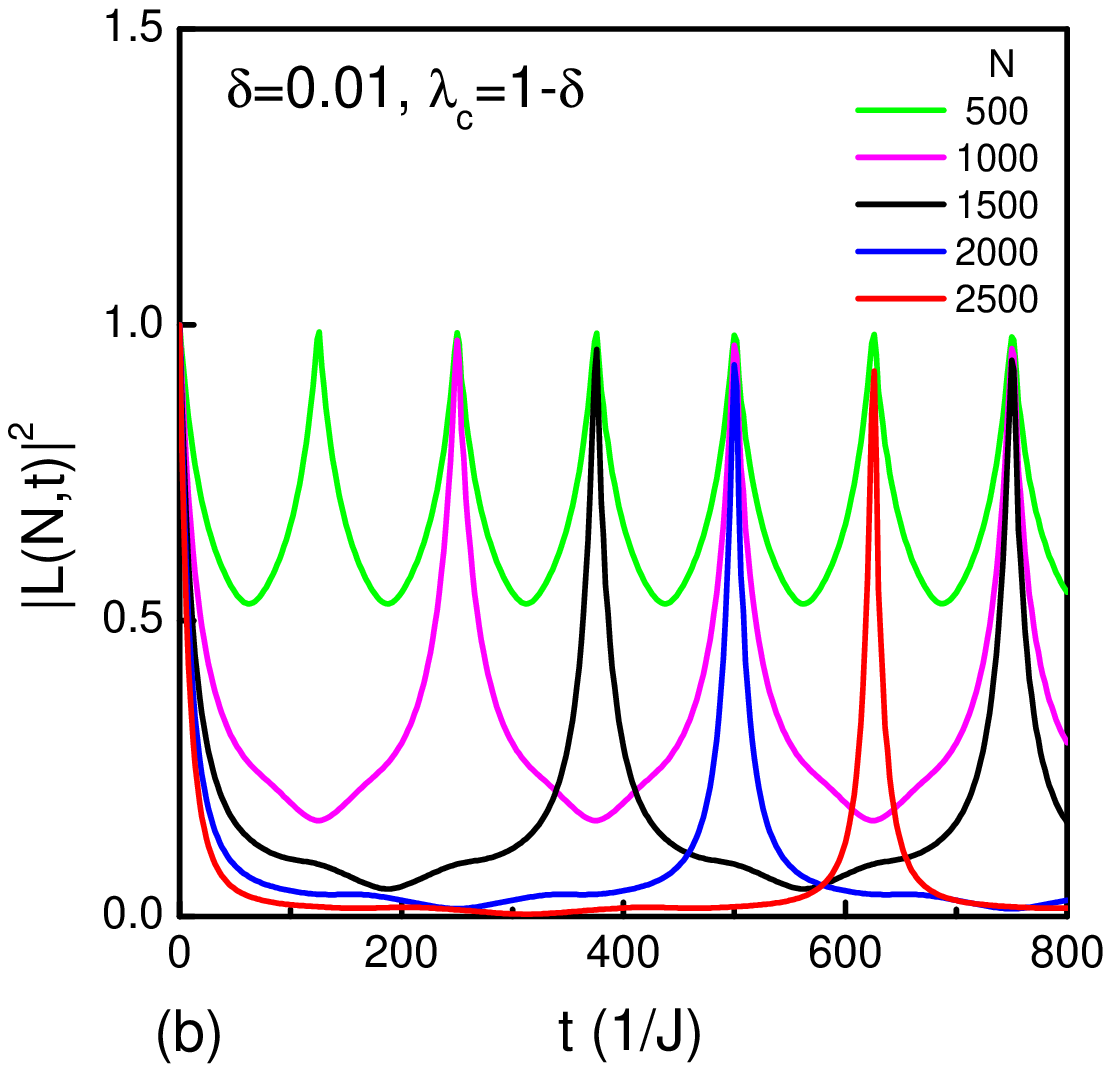}
\caption{(Color online) The quantum phase transition enhanced decay of LE at
large $N$ limit for small $\protect\delta $: except for $\protect\delta %
=0.01 $ the explanations are the same as that in Fig. 2.}
\end{figure}

Next, we numerically investigate the LE for system with finite $N$. It turns
out that as $N$ increases the LE will vanish for very small $\delta $. For
example, we take $\delta =0.01$, $N=500\sim 2500$ and compare the numerical
results illustrated in Fig. 3 with those for $\delta =0.1$, $N=50\sim 250$
in Fig. 2. From Fig. 2a and 3a one can clearly see that the valley narrows
as $\delta $ decreases and $N$ increases. On the other hand, the comparison
between Fig. 2b and 3b shows an interesting phenomenon in the LE at the
critical point $\lambda _{c}$. Firstly, the periods of the revival of the LE
for the two cases are both proportional to the size of $E$. Secondly, the
behaviors of the LE in Fig. 2b and 3b are almost same. The numerical results
seem to suggest an extrapolation for a scaling behavior, i.e., the LE at the
critical point $L_{c}(t,\delta ,N)$\ is invariant under the scaling
transformation $t\rightarrow t/\alpha $, $\delta \rightarrow \alpha \delta $%
, and $N\rightarrow N/\alpha $. This prediction is in agreement with the
analytical analysis with the small $Jt$\ approximations. Therefore, our
result opens a possibility that the distinctive picture of the decays and
revivals of the LE may serve as a good witness of QPT in the case of finite $%
N$.

\emph{Conclusion: }In summary, by a special model, we have obtained the
exact expression of the LE and analyzed the possible relation between the
quantum-classical transition of the central system, characterized by the LE
of the surrounding system, and the occurrence of a QPT in its surrounding
system $E$. Both the heuristic analysis and the numerical calculations we
performed reveal a novel mechanism of the decay of the LE. It is well known
that quantum critical phenomenon is closely associated with the entanglement
among the qubits consisting of the surrounding system \cite{QEQPT}. We would
like to stress that our present study is from a different perspective; our
emphasis is on the relation between the QPT of the surrounding system and
its entanglement with the central system $S$, which is qualitatively
characterized by the LE. In our model, the maximal quantum entanglement
between $S$ and $E$ can be reached and then the central system transits from
a pure state to a mixed state when a QPT of $E$ takes place in one of the
two evolution branches. This result suggests an unexplored and rather
intriguing relationship among entanglement, LE, decoherence and criticality.

This work is supported by the NSFC with grant Nos. 90203018, 10474104 and
60433050, and NFRPC with Nos. 2001CB309310 and 2005CB724508. CPS thanks T.
Xiang, L. Yu and Tien Kieu for helpful discussions.

\end{document}